\documentclass[letterpaper]{article} 
\usepackage{aaai25}  
\usepackage{times}  
\usepackage{helvet}  
\usepackage{courier}  
\usepackage[hyphens]{url}  
\usepackage{graphicx} 
\usepackage{natbib}  
\usepackage{caption} 
\usepackage{algorithm}
\usepackage{algorithmic}
\usepackage{amsmath}
\usepackage{multirow}
\usepackage{pifont}
\usepackage{graphicx}
\usepackage{booktabs}
\usepackage{arydshln}

\usepackage{xcolor}

\usepackage{newfloat}
\usepackage{listings}

\usepackage{newfloat}
\usepackage{listings}
\DeclareCaptionStyle{ruled}{labelfont=normalfont,labelsep=colon,strut=off} 
\floatstyle{ruled}
\newfloat{listing}{tb}{lst}{}
\floatname{listing}{Listing}
\title{Read and Think: An Efficient Step-wise Multimodal Language Model for
Document Understanding and Reasoning}
\author{
    Jinxu Zhang\textsuperscript{\rm 1} Qiyuan Fan\textsuperscript{\rm 1}  Yu Zhang \textsuperscript{\rm 1}}
\affiliations{
    \textsuperscript{\rm 1}Harbin Institute of Technology\\

    Harbin, Heilongjinag, China\\
    jxzhang@ir.hit.edu.cn
}

\usepackage{bibentry}

\begin{document}

\maketitle
\begin{abstract}
Understanding the contents of multimodal documents is essential to accurately extract relevant evidence and use it for reasoning. Existing document understanding models tend to generate answers with a single word or phrase directly, ignoring the source document's evidence and lacking interpretability. In this work, we address the lack of step-wise capabilities through data augmentation and extension. Specifically, We use Multi-modal Large Language Models (MLLMs), which have strong visual understanding and reasoning abilities, as data generators to generate step-wise question-and-answer pairs for document images and use a high-performance LLM as the error detector to filter out noisy data. This step-wise data generation pipeline is implemented using both template-based and few-shot methods. We then use the generated high-quality data to train a humanized document understanding and reasoning model, specifically designed to solve complex questions that require reasoning or multi-hop question answering, dubbed DocAssistant. Experimental results demonstrate the effectiveness and application value of step-wise generation, showing a 5\% improvement on InfoVQA with complex layouts and a 7\% improvement on ChartQA with complex reasoning, compared to directly generated answers. We hope our work highlights the potential of synthetic data and encourages further exploration of multi-modal document reasoning capabilities.
\end{abstract}

%

\section{Instruction}

\begin{figure}[h]
\centering
\captionsetup{font={small}}
\includegraphics[width=0.48\textwidth]{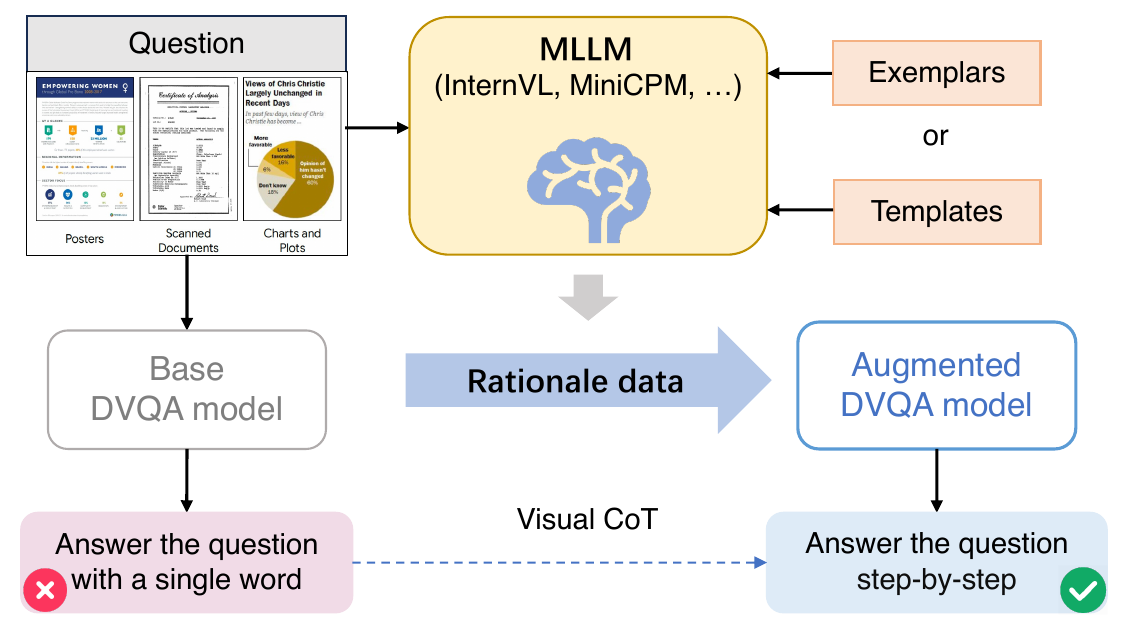}
\vspace{0.1em}
\caption{Existing document visual question answering models tend to generate a word or phrase directly as an answer, ignoring the evidence or reasoning steps of the source. We generate high-quality data with intermediate results using a large-scale multi-modal model by constructing templates and employing a few-shot approach. These augmented and extended data are then used to enhance a small-scale multi-modal model, achieving an efficient and general step-wise document understanding and reasoning model.}
\label{instruction}
\end{figure}

There are various documents in the real world, which differ from images of real-world scenarios. Document images are often filled with extensive text and graph information, requiring the model to have strong capabilities in document layout understanding, text semantic understanding, and numerical reasoning. This presents both a high challenge and high application value. In the document visual question answering (DVQA) task, as shown in Figure \ref{instruction}, existing document understanding models \cite{xu2020layoutlm,xu2020layoutlmv2,davis2022end} tend to generate answers directly based on questions, making it harder to determine the source of the answers and their accuracy. 
However, we aim for the model to think like a human, first locating the relevant information for questions in documents, and then extracting or inferring an answer based on that. For example, in text-rich documents, this would be the context related to keywords, while in chart documents \cite{methani2020plotqa,obeid2020chart,han2023chartllama}, it would be the relevant values and coordinates. This is a challenging task that requires the model to have both precise identification and strong reasoning abilities.

Recently, a number of MLLMs have made breakthroughs in document understanding, such as PaLI-3 \cite{chen2023pali}, InternVL \cite{chen2024internvl}, LlaVA-UHD \cite{xu2024llava}, and mPLUG-DocOwl \cite{hu2024mplug}. By improving image resolution and understanding documents from both macro and micro perspectives, these models have greatly enhanced the ability of visual encoders to understand fine-grained information in documents. Most of them have achieved promising performance on document understanding datasets like DocVQA \cite{mathew2021docvqa}, ChartQA \cite{masry2022chartqa}, InfographicVQA \cite{mathew2022infographicvqa}, etc. However, they can only handle simple questions in simple-layout documents, such as directly extracting answers from continuous text in the document images. When faced with complex-layout documents, they are likely to make errors in information identification. Additionally, when dealing with complex questions, they tend to ignore the reasoning steps and directly produce answers without providing the source or basis for these answers. Therefore, the existing models have a lot of room for improvement in the questions of documents with complex layouts and documents requiring reasoning.

At present, existing multi-modal document datasets are gradually improving, encompassing a range from the simple layout of scanned documents in DocVQA to the complex layout of poster documents in InfographicVQA, as well as chart documents involving mathematical logic operations in ChartQA. This progression allows for a more comprehensive assessment of the MLLMs' ability to understand different types of documents and various questions. However, most of the questions in these labeled datasets are relatively simple, focusing primarily on image information recognition, with fewer questions requiring reasoning. Additionally, the answers do not provide contextual information, making it difficult to intuitively confirm their correctness.

In this paper, We aim to develop an efficient and general document understanding model that combines document information recognition and reasoning. When answering questions, the model will think like a human: first locating the context information related to the keywords in the question, and then deducing the corresponding answer based on this context information. Since there is no such data at present, we enhance existing document visual question-and-answer data by taking the answer as the supervisory signal and allowing large-scale MLLMs to complete the thinking process from question to answer using our designed templates. Additionally, leveraging the powerful generative capabilities of MLLMs, we can extend the training data through the templates and samples we designed. Since the generated data contains noise, we designed a pipeline based on multi-agent collaborative filtering and rules to filter out these noisy data. Finally, the filtered data is used to train the small-scale vision language model (SLVM) to be efficient and proficient in document understanding and reasoning.

To summarize, our primary contributions include:
\begin{itemize}
    \item We used MLLMs to design a data generator based on templates and few-shots, and a multi-agents-based data filter to augment and extend high-quality, step-wise document visual question-and-answer data.
    \item We trained an efficient SLVM, dubbed DocAssistant, with both question-aware document visual context understanding and reasoning, which has a high practical value.
    \item We achieved state-of-the-art results on complex layout document understanding and reasoning datasets, and provided more extensive experiments to validate the superiority of our model.
\end{itemize}

\section{Related Work}

\subsection{Visually Rich Document Understanding (VRDU)}

The VRDU task is designed to provide a document image and a question, requiring the model to answer the question by understanding the text, images, and layout of the document. Existing document understanding models fall into two categories: OCR-based and end-to-end models that do not rely on external tools. The former includes models like LayoutLMv3 \cite{huang2022layoutlmv3}, UDOP \cite{tang2023unifying}, and DocFormerv2 \cite{appalaraju2024docformerv2}, which use OCR to extract text and corresponding coordinate information and design pre-training tasks based on text, layout, and image information to understand documents. The latter includes simpler architectures such as Donut \cite{kim2022ocr} and Pix2Struct \cite{lee2023pix2struct}, which use only a visual encoder and a text decoder to improve document understanding through pseudo-OCR tasks or pre-training tasks like image masking. With the rise of large language models, the multi-modal architecture model led by LLaVA \cite{liu2024visual} has also made significant breakthroughs. Models \cite{chen2023pali,hu2024mplug, chen2024far} involving document visual question-answering tasks have achieved good results in simple questions, such as document information extraction with simple layouts, by enhancing the visual encoder with the document data based on this architecture.

However, the above models do not perform well with documents that have a complex layout and questions requiring reasoning. To address this, we introduce a document visual chain of thought to train the SVLM by augmenting and extending the existing document datasets \cite{mathew2021docvqa, mathew2022infographicvqa, masry2022chartqa}. This allows the model to think like a human when answering questions, thereby improving the accuracy of its response.

\subsection{Reasoning step-by-step}

For LLMs, chain-of-thought prompting has been found to be a simple and effective method to improve reasoning performance, often used to tackle complex tasks. For example, SymbCoT \cite{xu2024faithful} uses LLMs to implement a system with translation, planning, solving, and verification, following a logical reasoning framework to maximize LLMs' ability to stimulate the chain of thought. In the multimodal field, recent works such as VisProg \cite{gupta2023visual} and ViperGPT \cite{suris2023vipergpt} utilize LLMs as planners to compose domain-specific models \cite{lu2024chameleon, yang2023mm, zeng2022socratic} to solve complex tasks.

However, in the field of document VQA, there is a lack of data and multimodal models capable of reasoning about document images. Given the powerful capabilities of LLMs in data generation, we employ a comprehensive approach that includes designing templates and exemplars, using tools, etc. This approach involves teaching a smaller model by learning from a larger teacher model to fill gaps in data and modeling for document visual question-and-answer tasks.

\section{Method}

\begin{figure*}[htbp]
    \centering
    \includegraphics[width=0.9\linewidth]{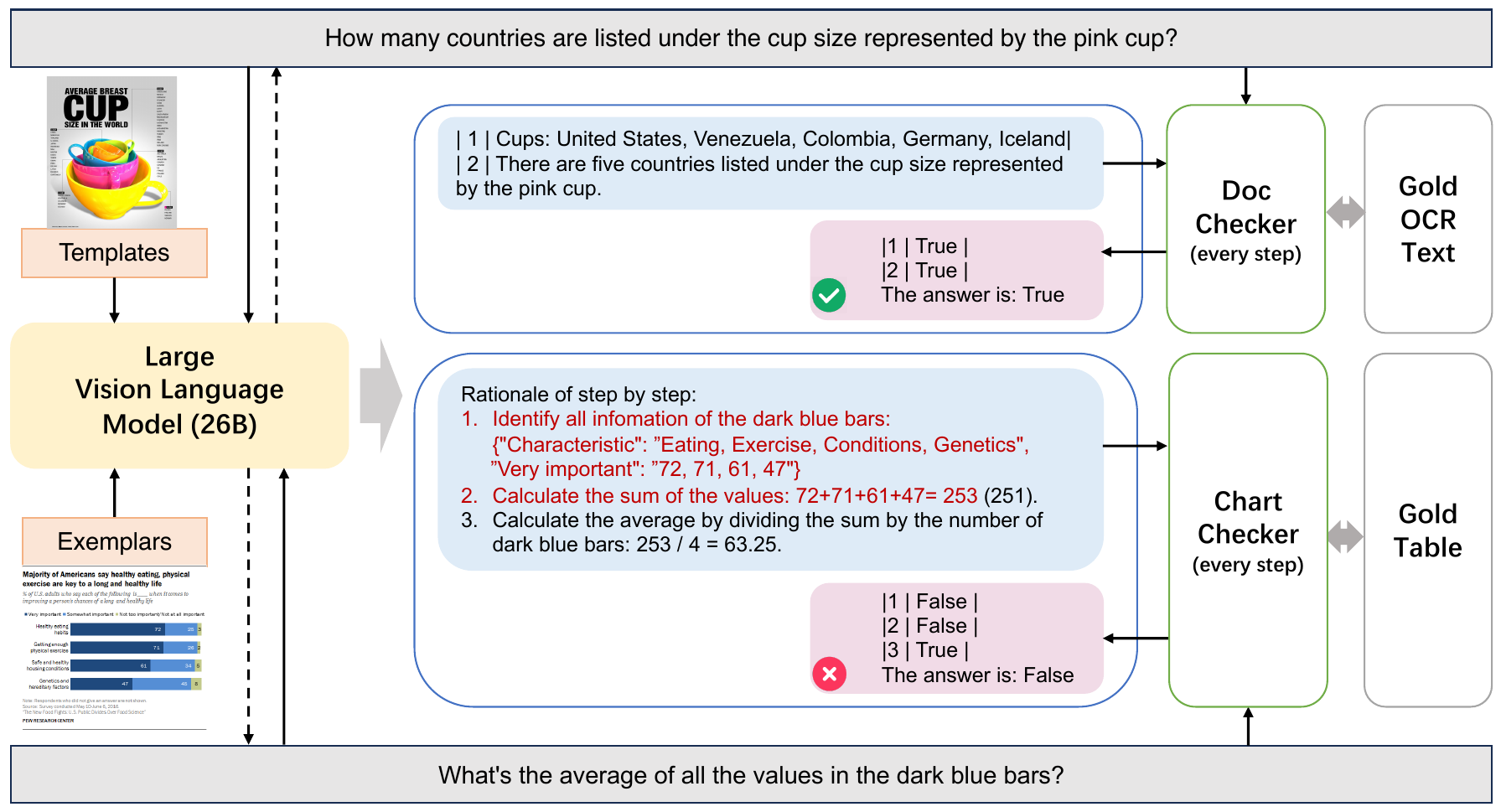} 
    \caption{Data checker based on multi-agent interaction. A 26B MLLM is used to generate answers and rationale with relevant context information for the corresponding text extracted document and chart reasoning document, according to the input templates or exemplars. Dotted arrows indicate the extended data including questions. The data checker uses OCR text from ordinary documents and tabular information equivalent to charts. First, it checks for extraction errors in the generated data. Second, it checks for errors in the intermediate steps of reasoning. If any errors are found, the data is considered unqualified.}
    \label{checker}
\end{figure*}

Given a document image and a question, the goal is to enable the model to think and answer step by step like a human. First, design a data generator and a data checker to construct high-quality data. These data are then used to train an efficient context-interpretable multi-modal document question-answering model.

In this section, We first describe the dataset construction process. Next, we explain how to filter out noisy data to retain high-quality data. Finally, we describe our model setup, using these data to train a dedicated multi-modal document question-answering model with contextual reasoning.

\subsection{Data Generator and Checker}

The overview of our MLLM-based data generator and checker is shown in Figure \ref{checker}. We leverage InternVL-1.5 \cite{chen2024far} as the MLLM for data generation and the LLaMA3-70B for data check. Data generation includes creating data based on the existing training set and generating triplets (question, rationale, answer) through our designed templates, which can be found in the Appendix. During the generation process based on the existing training set, we generate a related rationale $R$ using the template $P$, image $I$, question $Q$, and answer $A$ as inputs:
\begin{equation}
    R = |P;I;Q;A|
\end{equation}
During the triplet generation process, we design a question generation template, as shown in Table 4 of the Appendix. In particular, we use templates for extractive or abstractive documents and use few-shot learning for chart documents:
\begin{equation}
    (Q,R,A) = |P;I|
\end{equation}

During data inspection, we use tools to extract text information from documents, specifically using OCR tools for general documents and DePlot \cite{liu2022deplot} to convert charts into tables. By designing error detection templates shown in Table 7 in the Appendix, the LLM can determine whether the corresponding logical step is correct based on the question and the gold text of the image. This includes (1) the extraction of information step from the document itself and (2) the reasoning steps based on the extracted information. If any judgment errors are made, it will return False and delete the piece of data.

\subsection{Model Archecture}

\begin{figure*}
\includegraphics[width=0.9\textwidth]{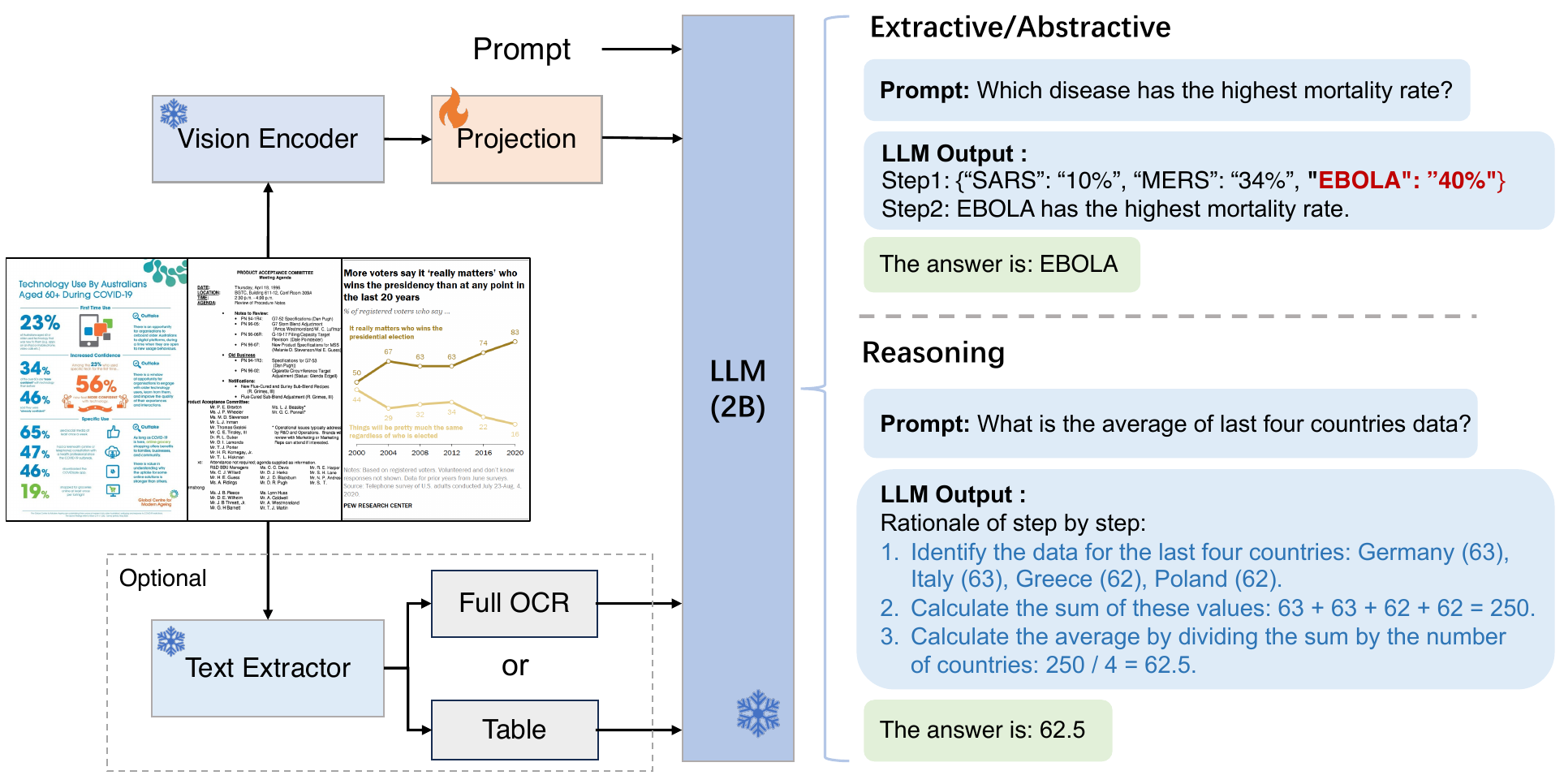}
\caption{Model overview. Document images, projected by projection layers concatenated with a prompt (question) and an optional OCR text or table, are fed into the language model for step-wise generation. Answer generation for the extractive and abstractive types consists of two steps: the first step generates the context relevant to the keywords in the question, and the second step generates the corresponding answer based on the context. For the reasoning type of answer generation, the steps depend on the question type and vary with the complexity of the question, using exemplars as the prompt.
}
\label{model}
\end{figure*}

The architecture of our trained model is shown in Figure \ref{model}. We leverage Mini-InternVL-Chat-2B as the backbone for DVQA. This model uses a pre-trained ViT \cite{dosovitskiy2020image} to extract image features, a trainable MLP module to map extracted image features to the text embedding space, and InternLM2-Chat-1.8B \cite{cai2024internlm2} as the language model to understand user instruction. 

Relying on pre-trained visual encoders alone to understand text-heavy images is insufficient. Therefore, external tools can be used to extract text information from the document as a supplement, and ensure that the context information understood by the model is accurate. 

\subsection{Extractive/Abstractive Answer Generation}

For the query document, the answer comes directly from certain parts of the document, and processing this answer generally requires two steps. First, it is necessary to locate the context information related to the keywords of the question, which may be one or more fragments, and then deduce the final answer based on this context information. That is, based on the keywords of the question (Q), the relevant contextual information fragments (C) are located in the document (D). Assuming we use the function \text{ExtractContext} to extract this context information, it can be expressed as:
\begin{equation}
 C = \text{ExtractContext}(D, Q) 
\end{equation}
where (C) represents the collection of contextual information fragments relevant to the question (Q). Next, based on the extracted context information (C), the final answer (A) is deduced. Suppose we use the function \text{InferAnswer} to infer the answer, then it can be expressed as:
\begin{equation}
 A = \text{InferAnswer}(C, Q) 
\end{equation}

\subsection{Reasoning Answer Generation}

Currently, existing reasoning document images mainly focus on numerical calculations, with the human dataset in ChartQA being a typical example. The answers are not directly present in the documents, they need to be deduced based on the data in the documents. The entire process includes multi-step calculations, where complex nested operations are decomposed to obtain more accurate answers. Since the number of steps is uncertain, we design few-shot templates during the training data generation process, enabling the model to respond flexibly to each question and generate corresponding reasoning steps. The whole process is as follows:

Given a document containing data (D), the question (Q), and the answer (A).
\begin{equation}
 D = \{d_1, d_2, \ldots, d_n\}
 \end{equation}
Flexibly respond to each question: design a few-shot template so that the model can flexibly generate reasoning steps for each question. Each step can be represented as:
\begin{equation}
 \text{steps} = \text{model}(\text{few\_shot\_prompt}) 
 \end{equation}
For each question (Q), the model generates corresponding reasoning steps until the final answer A is obtained.
\begin{equation}
 \text{steps} = \{\text{Step}_1, \text{Step}_2, \ldots, \text{Step}_m\}
 \end{equation}
 \begin{equation}
 A = \text{execute}(\text{steps}, \{d_i\}) 
 \end{equation}

\subsection{Training with rationales}

Due to the lack of data, especially since ChartQA's human training set has only 7,398 question pairs, we expanded the data in the original training set and developed a template for question generation to produce different types of data. We also filter noise data using our designed checker.

\section{Experiment}

\subsection{Experimental Setup}
\subsubsection{Dataset}
We run experiments on three document VQA datasets: DocVQA \cite{mathew2021docvqa}, InfographicVQA \cite{mathew2022infographicvqa}, and ChartQA \cite{masry2022chartqa}. DocVQA images come from scanned documents, including letters, tables, articles, etc. Most of the questions are relatively simple and involve text extraction directly. InfographicVQA images are taken from posters, where the layout is complex. Most of the questions involve text extraction, and some require simple reasoning. The charts in the ChartQA dataset include bar charts, line charts, and pie charts, all of which are synthetic data or human-marked data. Most of the synthetic data are simple chart information identification, while the human-marked data contain complex mathematical logic operations.

\subsubsection{Generated Data}
Table \ref{filter} shows the generated QA statistics. The sum includes data based on the original training set and data generated from the question generation template extension. The template can be found in the Appendix (Table 3,4,5). For DocVQA, 3 new questions were generated for each image, while InfographicVQA had 4 new questions per image. Given the small amount of original data in ChartQA Human, Since there is less raw data in ChartQA Human, we collected the related Chart images of Chart-to-text \cite{kantharaj-etal-2022-chart} and the corresponding gold table to generate more chart reasoning data similar to human. Filtered indicates the data filtered through the LLM and rules.

In addition, we generated questions based on the types of questions in each dataset, such as Count, Spatial, and Reasoning data for DocVQA. For more details, see the Appendix (Table 6 and Table 8). 

\begin{table}[]
\centering
\begin{tabular}{ccccc}
\specialrule{1.5pt}{1.5pt}{1.5pt}
Dataset        & Images & \multicolumn{2}{c}{Generated} & Filtered \\ \specialrule{1.5pt}{1.5pt}{1.5pt}
DocVQA         & 10194  & \multicolumn{2}{c}{39459+30582}     & 58324    \\
InfoVQA        & 4406   & \multicolumn{2}{c}{23945+17624}     & 36832    \\
ChartQA        & 18317+44096  & \multicolumn{2}{c}{7398+80831}     & 67649    \\ \specialrule{1.5pt}{1.5pt}{1.5pt}
\end{tabular}
\caption{Statistics of generated data.}
\label{filter}
\end{table}

\subsubsection{Compared Methods}
We compare existing document understanding models across various categories: (1) plain text models represented by T5, (2) the LayoutLM series represented by LayoutLMv3, and DocFormerv2, which has the best performance among the T+L+V models, (3) the first OCR-free model Donut for understanding documents with image encoders, and Pix2Struct for performing best in end-to-end small-scale document understanding models, and (4) multimodal document understanding models combined with LLMs.

\subsubsection{Implementation Details} 

During the training process, we set 2 epochs with a batch size of 8 and a learning rate of 4e-5. Specifically, dynamically resizing the image, adjusting the resolution to 448x448, and setting the maximum patch size to 12 are crucial for understanding document-type images. All experiments were conducted on a 2B model, with training focused solely on the projection layer and the language model incorporating LoRA.

\subsection{Main Results}

\begin{table*}[]
\centering
\begin{tabular}{cccclclcl}
\specialrule{1.5pt}{1.5pt}{1.5pt}
\multirow{2}{*}{Method} & \multirow{2}{*}{Modality} & \multirow{2}{*}{Param}   & \multicolumn{2}{c}{DocVQA} & \multicolumn{2}{c}{InfographicVQA} & \multicolumn{2}{c}{ChartQA}        \\ \cline{4-9} 
                        &                           &                          & \multicolumn{2}{c}{ANLS}   & \multicolumn{2}{c}{ANLS}           & \multicolumn{2}{c}{Relax Accuracy} \\ \specialrule{1.5pt}{1.5pt}{1.5pt}
T5 \cite{2020t5}                      & T                         & 223M                     & \multicolumn{2}{c}{70.4}   & \multicolumn{2}{c}{36.7}           & \multicolumn{2}{c}{59.8}           \\ \hline
LayoutLMv3 \cite{huang2022layoutlmv3}              & \multicolumn{1}{l}{T+V+L} & \multicolumn{1}{l}{368M} & \multicolumn{2}{c}{83.37}  & \multicolumn{2}{c}{45.1}           & \multicolumn{2}{c}{-}              \\
DocFormerv2 \cite{appalaraju2024docformerv2}             & \multicolumn{1}{l}{T+V+L} & \multicolumn{1}{l}{750M} & \multicolumn{2}{c}{\underline{87.84}}  & \multicolumn{2}{c}{48.8}           & \multicolumn{2}{c}{-}              \\ \hline
Donut \cite{kim2022ocr}                   & V                         & 201M                     & \multicolumn{2}{c}{67.5}   & \multicolumn{2}{c}{11.5}           & \multicolumn{2}{c}{41.8}           \\
Pix2Struct \cite{lee2023pix2struct}              & V                         & 1.3B                     & \multicolumn{2}{c}{76.6}   & \multicolumn{2}{c}{40.0}           & \multicolumn{2}{c}{58.6}           \\ \hline
UReader \cite{ye2023ureader}                 & V                         & 7B                       & \multicolumn{2}{c}{65.4}   & \multicolumn{2}{c}{42.2}           & \multicolumn{2}{c}{59.3}           \\
Qwen-VL \cite{bai2023qwen}                 & V                         & 7B                       & \multicolumn{2}{c}{65.1}   & \multicolumn{2}{c}{29.9}           & \multicolumn{2}{c}{65.7}           \\
mPLUG-DocOwl \cite{hu2024mplug}              & V                         & 8B                       & \multicolumn{2}{c}{62.2}   & \multicolumn{2}{c}{38.2}           & \multicolumn{2}{c}{57.4}           \\ 
SMoLA-PaLI-3 \cite{Wu_2024_CVPR}              & V                         & 5B                       & 
\multicolumn{2}{c}{84.5}   & \multicolumn{2}{c}{\underline{52.4}}           & \multicolumn{2}{c}{\underline{68.9}}           \\  \hdashline
DocAssistant (Ours)     & V                         & 2B                       & \multicolumn{2}{c}{85.58}       & \multicolumn{2}{c}{\textbf{62.45}}               & \multicolumn{2}{c}{\textbf{77.44}}               \\ \hline
With gold text pipeline \\ \hline\hline
SMoLA-PaLI-3 \cite{Wu_2024_CVPR}              & T+V                         & 5B                       & 
\multicolumn{2}{c}{87.4}   & \multicolumn{2}{c}{57.3}           & \multicolumn{2}{c}{68.9}           \\ 

DocAssistant(Ours)         & T+V                       & 2B                       & \multicolumn{2}{c}{\textbf{88.12}}       & \multicolumn{2}{c}{\textbf{65.72}}               & \multicolumn{2}{c}{\textbf{78.28}}               \\ \specialrule{1.5pt}{1.5pt}{1.5pt}
\end{tabular}
\caption{Comparison with models of different scales and different modal models. Underlined values represent the SOTA of previous methods. T, L, and V represent text, layout, and visual information, respectively. The evaluation method of DocVQA and InfoVQA is ANLS, while the evaluation method of ChartQA is Relax Accuracy.}
\label{main_results}
\end{table*}

\subsubsection{Comparison with Existing Models}
Existing models only provide a single word or phrase as an answer for the DVQA task. On one hand, it is impossible to determine its source. On the other hand, for complex questions requiring reasoning, ignoring intermediate reasoning steps is more likely to lead to errors. Through our reconstruction of the dataset and efficient training on the 2B multimodal model, we have achieved impressive results, as shown in Table \ref{main_results}. Using only a small number of parameters, it outperforms many larger document understanding models and demonstrates greater reasoning power than other existing models.

In terms of details, DocAssistant's performance on DocVQA is inferior to DocFormerv2, indicating that understanding multimodal information of documents is more accurate than understanding only image information. However, existing models perform poorly for documents with complex layouts and documents with complex questions requiring reasoning like InfoVQA and ChartQA. In particular, small-scale models have little or no document reasoning ability. We have addressed this gap while surpassing larger models ($\leq$ 8B) in document understanding and reasoning.

Additionally, due to the large amount of text information contained in the document, it is difficult for the visual encoder to fully comprehend all the fine-grained semantic content. External tools can be used to extract textual information as a supplement. The experimental results in Table \ref{main_results} show that the model achieved improvements on the three datasets by incorporating an understanding of multi-modal document information, surpassing existing methods.

\subsubsection{Comparison of Different Strategies}
\begin{table}[h]
\centering
\begin{tabular}{ccccc}
\specialrule{1.5pt}{1.5pt}{1.5pt}
\multirow{2}{*}{Strategy} & \multirow{2}{*}{DocVQA} & \multirow{2}{*}{InfoVQA} & \multicolumn{2}{c}{ChartQA} \\ \cline{4-5} 
                          &                         &                          & aug.         & human        \\ \hline
single-word               & 81.02                   & 49.38                    & 68.96        & 51.04        \\ \hline
zero-shot                 & 79.58                   & 53.43                    & 72.8         & 52.88        \\
few-shot                  & 78.13                   & 51.09                    & 77.68        & 54.96        \\
finetune                  & \textbf{81.83}                      & \textbf{54.64}                         & \textbf{81.28}        & \textbf{62.32}        \\  \specialrule{1.5pt}{1.5pt}{1.5pt}
\end{tabular}
\caption{The comparison results of DocAssistant under different strategies, and the evaluation method is Accuracy.}
\label{strategy}
\end{table}

In the experiment of strategy in Table \ref{strategy}, the zero-shot setting involved changing the original prompt to "Answer the question step by step". According to the experimental results, the model performs better on complex layout documents and complex reasoning questions when answering questions step by step. For extracted answers to simple documents like DocVQA, the results are similar. For the few-shot setup, we used 3-shot in the experiment and found that the instruction compliance of the 2B model was poor. Its answers did not follow the example format, especially in DocVQA and InfoVQA, which did not require reasoning and had a casual answer style. Although ChartQA did not follow the example format, it still answered step by step, leading to improved performance, while the other two datasets with extracted answers performed poorly. Since generation styles of different strategies are different, accuracy is used as the evaluation metric.

Single-word answers can more directly fit the answer, but they lack contextual information, are less interpretable, and are more difficult to judge the correctness. Zero-shot and few-shot are unstable and heavily influenced by instruction settings. For documents with extractive or abstractive answers, most answers still do not provide source information. These are the original intention behind our approach to building step-by-step datasets and training the model. The results of fine-tuning show that the model's reasoning ability has greatly improved. Not only is the answer accuracy higher but the source information of the answer can also be provided, helping users confirm the reliability of the answer.

\subsection{Ablations and Analysis}

\begin{table}[h] 
\resizebox{0.47\textwidth}{!}{ 
\begin{tabular}{cccc}
\specialrule{1.5pt}{1.5pt}{1.5pt}
              & DocVQA & InfoVQA & ChartQA \\ \hline
before\_extend & 83.15       & 59.64        & 72.32        \\
after\_extend  & 84.48       & 60.87        & 74.25        \\
after\_check  & 85.58       &  62.45       & 77.44        \\ \specialrule{1.5pt}{1.5pt}{1.5pt}
\end{tabular}
}
\caption{Before and after using extended data and before and after using filtered training data. The evaluation method of DocVQA and InfoVQA is ANLS, while the evaluation method of ChartQA is Relax Accuracy.}
\label{extend}
\end{table}
\subsubsection{The Effective of Extend Data}
The modification of the original training set data, combined with our extended data, is sufficient for the model to learn the required response style. Experimental results in Table \ref{extend} show a performance gain after extension. This also provides a data generation tool for other smaller models, enabling them to further improve performance by generating more data.
\subsubsection{The Effective of Checker} 
We fine-tuned the model using both pre-filtered and filtered data and the results in Table \ref{extend} showed improved performance on all three datasets, especially InfoVQA and ChartQA. This indicates that noisy data significantly interfered with documents with complex layouts and proves the effectiveness of our designed checker.

\subsubsection{Qualitative Analysis}

\begin{figure*}
\centering
\includegraphics[width=0.9\textwidth]{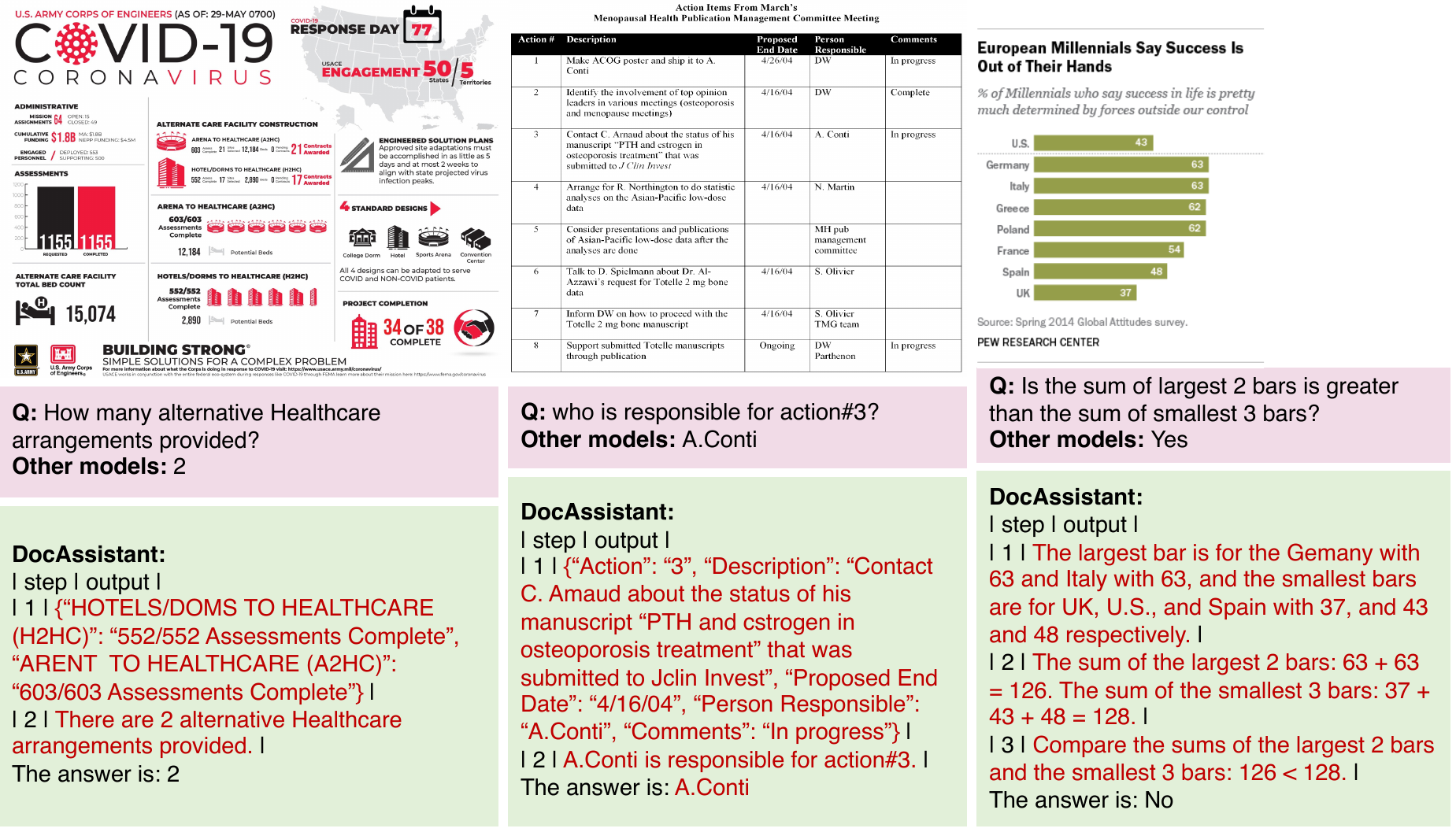}
\caption{Output comparison of DocAssistant and other models on three datasets, with red font representing rationales relevant to the question.
}
\label{quality}
\end{figure*}

Figure \ref{quality} shows a comparison between cases generated by DocAssistant and those generated by other models (T, T+L+V, V). The answers provided by other models in different documents are all single words or phrases, which tend to be incorrect when faced with questions such as chart reasoning and document information extraction with complex layouts, and they do not contain contextual information. The first two examples are questions and answers about complex layout documents, where DocAssistant provides detailed context. The third example is a chart reasoning question, where DocAssistant also gives specific reasoning steps, providing stronger interpretation and higher accuracy and allowing users to understand how the model reached the answer and easily identify any mistakes. Most importantly, model performance can be further improved through extensive training with step-wise data.

However, due to the model's inherent limitations, DocAssistant still makes mistakes in reasoning on complex questions. Related error case studies can be found in the Appendix Figure 1 and Figure 2.

\subsubsection{Analysis of Different Question Types}

\begin{figure}
\centering
\includegraphics[width=0.47\textwidth]{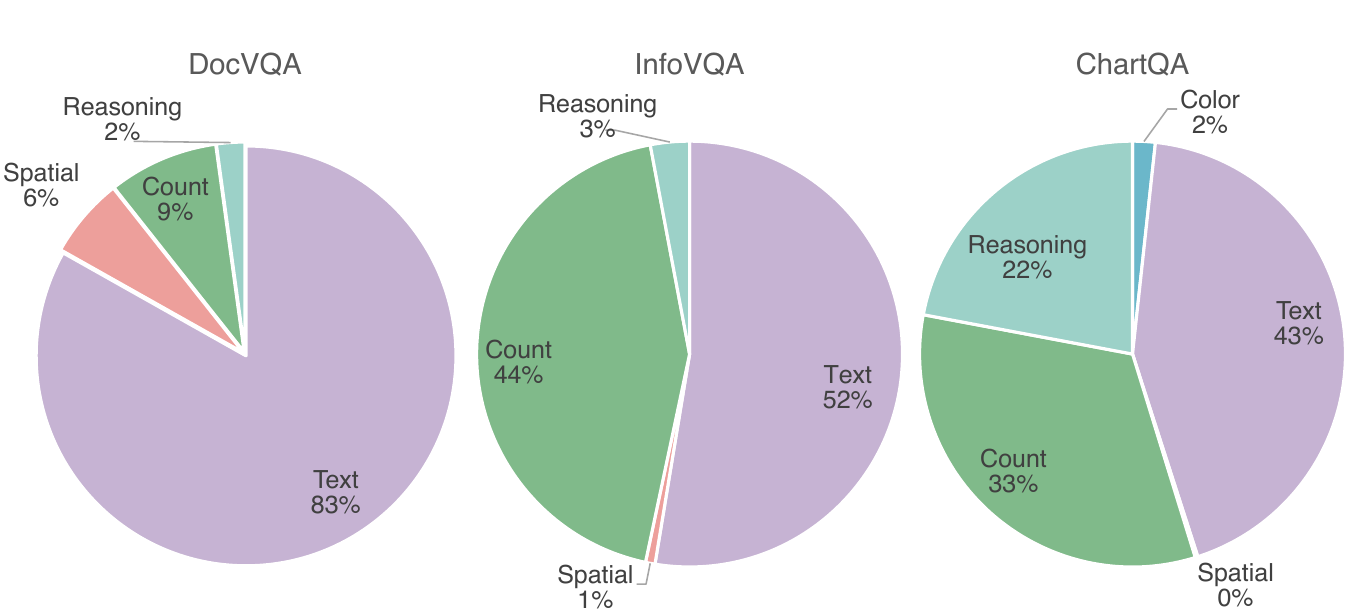}
\caption{Analysis of different types of questions in three test sets.
}
\label{count}
\end{figure}

Figure \ref{count} shows the proportion of question types of the three datasets. The Text\_extractive type has the largest proportion in all three datasets, especially in DocVQA. InfoVQA and ChartQA have a higher proportion of Count and Reasoning data, respectively, where DocAssistant shows more advantages.”

Table \ref{type_result} shows how five types of questions in three different types of datasets are represented in different strategies. The experimental results show that the SFT strategy (DocAssistant) has the best performance for all count and reasoning questions in the three datasets. In addition to DocVQA (simple layout, simple question), zero-shot and few-shot performed better than the straightforward setup on InfoVQA (complex layout) and ChartQA (complex question), indicating that the chain-of-thought is effective in solving questions of such documents. It is worth noting that due to the small amount of color and spatial data, there is a large variance in different method settings. For example, ChartQA only has 43 Color-type questions and 5 Spatial-type questions. Interestingly, both few-shot and step-wise training can reduce the comprehension of color information, The possible reason is that the visual encoder's recognition performance of color information is unstable.

\begin{table}[h]
\centering
\setlength{\tabcolsep}{2pt}
\begin{tabular}{ccccccc}
\specialrule{1.5pt}{1.5pt}{1.5pt}
Dataset                  & Method    & Color & Text & Spatial & Count & Reasoning \\ \hline
\multirow{4}{*}{DocVQA}  & SF       &-       &81.65      &85.46         &80.04       &51.69           \\
                        & ZS       &-       &79.77      &83.48         &81.0       &61.48           \\
                        & FS       &-       &78.33      &79.24         &81.43       &55.28           \\
                         & SFT &-       &82.03      &84.53         &81.24       &65.72           \\ \hline
\multirow{4}{*}{InfoVQA} & SF        &44.83       &50.32      &28.57         &49.13       &43.37           \\
                        & ZS       &55.17       &48.65      &30.0         &58.85       &74.12           \\
                        & FS       &51.72       &47.82      &25.0         &54.04       &70.24           \\
                         & SW &58.62       &48.76      &30.0         &59.84       &77.01           \\ \hline
\multirow{4}{*}{ChartQA} & SF        &67.44       &76.05      &75.0         &52.50       & 50.74          \\
                        & ZS       & 65.12      &77.90      & 87.50        &54.37       &55.09           \\
                        & FS       &32.56       &79.17      &12.5         & 60.45      &51.29           \\
                         & SFT &39.53       &78.80      & 1.0        &71.65       &70.61           \\ \specialrule{1.5pt}{1.5pt}{1.5pt}
\end{tabular}
\caption{Comparison of different strategies on different types of questions in all datasets. Here, SF (Straightforward) stands for generating a word or phrase directly, ZS stands for Zero-shot, FS represents Few-shot, and SFT indicates a step-wise result trained not with a special prompt but with the generated data. The evaluation also uses Accuracy.}
\label{type_result}
\end{table}

\section{Conclusion}
This paper introduces a new model named DocAssistant, designed to perform document understanding and reasoning. Specifically, we first transform the data of the existing document training set to include the intermediate analysis process in answering questions and then expand the training data by adding missing types such as Spatial, Count, and Reasoning, a data checker is designed during this period for generated data, resulting in high-quality data that can then be trained using an efficient 2B model. The results surpass existing larger-scale models and are state-of-the-art across three datasets. Additionally, we conducted extensive experiments, comparing different strategies of the model, performance before and after data expansion and filtering, and various types of questions to validate our work.

However, the current general document visual question answering data for reasoning is still scarce, and SVLMs also have shortcomings in understanding document elements such as color and spatial information. In the future, we aim to address these deficiencies and hope our work will contribute to advancements in the field.
\bibliography{LaTeX/aaai25}

\begin{thebibliography}{34}
\providecommand{\natexlab}[1]{#1}

\bibitem[{Appalaraju et~al.(2024)Appalaraju, Tang, Dong, Sankaran, Zhou, and Manmatha}]{appalaraju2024docformerv2}
Appalaraju, S.; Tang, P.; Dong, Q.; Sankaran, N.; Zhou, Y.; and Manmatha, R. 2024.
\newblock Docformerv2: Local features for document understanding.
\newblock In \emph{Proceedings of the AAAI Conference on Artificial Intelligence}, volume~38, 709--718.

\bibitem[{Bai et~al.(2023)Bai, Bai, Yang, Wang, Tan, Wang, Lin, Zhou, and Zhou}]{bai2023qwen}
Bai, J.; Bai, S.; Yang, S.; Wang, S.; Tan, S.; Wang, P.; Lin, J.; Zhou, C.; and Zhou, J. 2023.
\newblock Qwen-vl: A frontier large vision-language model with versatile abilities.
\newblock \emph{arXiv preprint arXiv:2308.12966}.

\bibitem[{Chen et~al.(2023)Chen, Wang, Beyer, Kolesnikov, Wu, Voigtlaender, Mustafa, Goodman, Alabdulmohsin, Padlewski et~al.}]{chen2023pali}
Chen, X.; Wang, X.; Beyer, L.; Kolesnikov, A.; Wu, J.; Voigtlaender, P.; Mustafa, B.; Goodman, S.; Alabdulmohsin, I.; Padlewski, P.; et~al. 2023.
\newblock Pali-3 vision language models: Smaller, faster, stronger.
\newblock \emph{arXiv preprint arXiv:2310.09199}.

\bibitem[{Chen et~al.(2024{\natexlab{a}})Chen, Wang, Tian, Ye, Gao, Cui, Tong, Hu, Luo, Ma et~al.}]{chen2024far}
Chen, Z.; Wang, W.; Tian, H.; Ye, S.; Gao, Z.; Cui, E.; Tong, W.; Hu, K.; Luo, J.; Ma, Z.; et~al. 2024{\natexlab{a}}.
\newblock How far are we to gpt-4v? closing the gap to commercial multimodal models with open-source suites.
\newblock \emph{arXiv preprint arXiv:2404.16821}.

\bibitem[{Chen et~al.(2024{\natexlab{b}})Chen, Wu, Wang, Su, Chen, Xing, Zhong, Zhang, Zhu, Lu et~al.}]{chen2024internvl}
Chen, Z.; Wu, J.; Wang, W.; Su, W.; Chen, G.; Xing, S.; Zhong, M.; Zhang, Q.; Zhu, X.; Lu, L.; et~al. 2024{\natexlab{b}}.
\newblock Internvl: Scaling up vision foundation models and aligning for generic visual-linguistic tasks.
\newblock In \emph{Proceedings of the IEEE/CVF Conference on Computer Vision and Pattern Recognition}, 24185--24198.

\bibitem[{Davis et~al.(2022)Davis, Morse, Price, Tensmeyer, Wigington, and Morariu}]{davis2022end}
Davis, B.; Morse, B.; Price, B.; Tensmeyer, C.; Wigington, C.; and Morariu, V. 2022.
\newblock End-to-end document recognition and understanding with dessurt.
\newblock In \emph{European Conference on Computer Vision}, 280--296. Springer.

\bibitem[{Dosovitskiy et~al.(2020)Dosovitskiy, Beyer, Kolesnikov, Weissenborn, Zhai, Unterthiner, Dehghani, Minderer, Heigold, Gelly et~al.}]{dosovitskiy2020image}
Dosovitskiy, A.; Beyer, L.; Kolesnikov, A.; Weissenborn, D.; Zhai, X.; Unterthiner, T.; Dehghani, M.; Minderer, M.; Heigold, G.; Gelly, S.; et~al. 2020.
\newblock An image is worth 16x16 words: Transformers for image recognition at scale.
\newblock \emph{arXiv preprint arXiv:2010.11929}.

\bibitem[{Gupta and Kembhavi(2023)}]{gupta2023visual}
Gupta, T.; and Kembhavi, A. 2023.
\newblock Visual programming: Compositional visual reasoning without training.
\newblock In \emph{Proceedings of the IEEE/CVF Conference on Computer Vision and Pattern Recognition}, 14953--14962.

\bibitem[{Han et~al.(2023)Han, Zhang, Chen, Yang, Wang, Yu, Fu, and Zhang}]{han2023chartllama}
Han, Y.; Zhang, C.; Chen, X.; Yang, X.; Wang, Z.; Yu, G.; Fu, B.; and Zhang, H. 2023.
\newblock Chartllama: A multimodal llm for chart understanding and generation.
\newblock \emph{arXiv preprint arXiv:2311.16483}.

\bibitem[{Hu et~al.(2024)Hu, Xu, Ye, Yan, Zhang, Zhang, Li, Zhang, Jin, Huang et~al.}]{hu2024mplug}
Hu, A.; Xu, H.; Ye, J.; Yan, M.; Zhang, L.; Zhang, B.; Li, C.; Zhang, J.; Jin, Q.; Huang, F.; et~al. 2024.
\newblock mPLUG-DocOwl 1.5: Unified Structure Learning for OCR-free Document Understanding.
\newblock \emph{arXiv preprint arXiv:2403.12895}.

\bibitem[{Huang et~al.(2022)Huang, Lv, Cui, Lu, and Wei}]{huang2022layoutlmv3}
Huang, Y.; Lv, T.; Cui, L.; Lu, Y.; and Wei, F. 2022.
\newblock Layoutlmv3: Pre-training for document ai with unified text and image masking.
\newblock In \emph{Proceedings of the 30th ACM International Conference on Multimedia}, 4083--4091.

\bibitem[{Kantharaj et~al.(2022)Kantharaj, Leong, Lin, Masry, Thakkar, Hoque, and Joty}]{kantharaj-etal-2022-chart}
Kantharaj, S.; Leong, R.~T.; Lin, X.; Masry, A.; Thakkar, M.; Hoque, E.; and Joty, S. 2022.
\newblock Chart-to-Text: A Large-Scale Benchmark for Chart Summarization.
\newblock In Muresan, S.; Nakov, P.; and Villavicencio, A., eds., \emph{Proceedings of the 60th Annual Meeting of the Association for Computational Linguistics (Volume 1: Long Papers)}, 4005--4023. Dublin, Ireland: Association for Computational Linguistics.

\bibitem[{Kim et~al.(2022)Kim, Hong, Yim, Nam, Park, Yim, Hwang, Yun, Han, and Park}]{kim2022ocr}
Kim, G.; Hong, T.; Yim, M.; Nam, J.; Park, J.; Yim, J.; Hwang, W.; Yun, S.; Han, D.; and Park, S. 2022.
\newblock Ocr-free document understanding transformer.
\newblock In \emph{European Conference on Computer Vision}, 498--517. Springer.

\bibitem[{Lee et~al.(2023)Lee, Joshi, Turc, Hu, Liu, Eisenschlos, Khandelwal, Shaw, Chang, and Toutanova}]{lee2023pix2struct}
Lee, K.; Joshi, M.; Turc, I.~R.; Hu, H.; Liu, F.; Eisenschlos, J.~M.; Khandelwal, U.; Shaw, P.; Chang, M.-W.; and Toutanova, K. 2023.
\newblock Pix2struct: Screenshot parsing as pretraining for visual language understanding.
\newblock In \emph{International Conference on Machine Learning}, 18893--18912. PMLR.

\bibitem[{Liu et~al.(2022)Liu, Eisenschlos, Piccinno, Krichene, Pang, Lee, Joshi, Chen, Collier, and Altun}]{liu2022deplot}
Liu, F.; Eisenschlos, J.~M.; Piccinno, F.; Krichene, S.; Pang, C.; Lee, K.; Joshi, M.; Chen, W.; Collier, N.; and Altun, Y. 2022.
\newblock Deplot: One-shot visual language reasoning by plot-to-table translation.
\newblock \emph{arXiv preprint arXiv:2212.10505}.

\bibitem[{Liu et~al.(2024)Liu, Li, Wu, and Lee}]{liu2024visual}
Liu, H.; Li, C.; Wu, Q.; and Lee, Y.~J. 2024.
\newblock Visual instruction tuning.
\newblock \emph{Advances in neural information processing systems}, 36.

\bibitem[{Lu et~al.(2024)Lu, Peng, Cheng, Galley, Chang, Wu, Zhu, and Gao}]{lu2024chameleon}
Lu, P.; Peng, B.; Cheng, H.; Galley, M.; Chang, K.-W.; Wu, Y.~N.; Zhu, S.-C.; and Gao, J. 2024.
\newblock Chameleon: Plug-and-play compositional reasoning with large language models.
\newblock \emph{Advances in Neural Information Processing Systems}, 36.

\bibitem[{Masry et~al.(2022)Masry, Long, Tan, Joty, and Hoque}]{masry2022chartqa}
Masry, A.; Long, D.~X.; Tan, J.~Q.; Joty, S.; and Hoque, E. 2022.
\newblock Chartqa: A benchmark for question answering about charts with visual and logical reasoning.
\newblock \emph{arXiv preprint arXiv:2203.10244}.

\bibitem[{Mathew et~al.(2022)Mathew, Bagal, Tito, Karatzas, Valveny, and Jawahar}]{mathew2022infographicvqa}
Mathew, M.; Bagal, V.; Tito, R.; Karatzas, D.; Valveny, E.; and Jawahar, C. 2022.
\newblock Infographicvqa.
\newblock In \emph{Proceedings of the IEEE/CVF Winter Conference on Applications of Computer Vision}, 1697--1706.

\bibitem[{Mathew, Karatzas, and Jawahar(2021)}]{mathew2021docvqa}
Mathew, M.; Karatzas, D.; and Jawahar, C. 2021.
\newblock Docvqa: A dataset for vqa on document images.
\newblock In \emph{Proceedings of the IEEE/CVF winter conference on applications of computer vision}, 2200--2209.

\bibitem[{Methani et~al.(2020)Methani, Ganguly, Khapra, and Kumar}]{methani2020plotqa}
Methani, N.; Ganguly, P.; Khapra, M.~M.; and Kumar, P. 2020.
\newblock Plotqa: Reasoning over scientific plots.
\newblock In \emph{Proceedings of the IEEE/CVF Winter Conference on Applications of Computer Vision}, 1527--1536.

\bibitem[{Obeid and Hoque(2020)}]{obeid2020chart}
Obeid, J.; and Hoque, E. 2020.
\newblock Chart-to-text: Generating natural language descriptions for charts by adapting the transformer model.
\newblock \emph{arXiv preprint arXiv:2010.09142}.

\bibitem[{Raffel et~al.(2020)Raffel, Shazeer, Roberts, Lee, Narang, Matena, Zhou, Li, and Liu}]{2020t5}
Raffel, C.; Shazeer, N.; Roberts, A.; Lee, K.; Narang, S.; Matena, M.; Zhou, Y.; Li, W.; and Liu, P.~J. 2020.
\newblock Exploring the Limits of Transfer Learning with a Unified Text-to-Text Transformer.
\newblock \emph{Journal of Machine Learning Research}, 21(140): 1--67.

\bibitem[{Sur{\'\i}s, Menon, and Vondrick(2023)}]{suris2023vipergpt}
Sur{\'\i}s, D.; Menon, S.; and Vondrick, C. 2023.
\newblock Vipergpt: Visual inference via python execution for reasoning.
\newblock In \emph{Proceedings of the IEEE/CVF International Conference on Computer Vision}, 11888--11898.

\bibitem[{Tang et~al.(2023)Tang, Yang, Wang, Fang, Liu, Zhu, Zeng, Zhang, and Bansal}]{tang2023unifying}
Tang, Z.; Yang, Z.; Wang, G.; Fang, Y.; Liu, Y.; Zhu, C.; Zeng, M.; Zhang, C.; and Bansal, M. 2023.
\newblock Unifying vision, text, and layout for universal document processing.
\newblock In \emph{Proceedings of the IEEE/CVF Conference on Computer Vision and Pattern Recognition}, 19254--19264.

\bibitem[{Wu et~al.(2024)Wu, Hu, Wang, Pang, and Soricut}]{Wu_2024_CVPR}
Wu, J.; Hu, X.; Wang, Y.; Pang, B.; and Soricut, R. 2024.
\newblock Omni-SMoLA: Boosting Generalist Multimodal Models with Soft Mixture of Low-rank Experts.
\newblock In \emph{Proceedings of the IEEE/CVF Conference on Computer Vision and Pattern Recognition (CVPR)}, 14205--14215.

\bibitem[{Xu et~al.(2024{\natexlab{a}})Xu, Fei, Pan, Liu, Lee, and Hsu}]{xu2024faithful}
Xu, J.; Fei, H.; Pan, L.; Liu, Q.; Lee, M.-L.; and Hsu, W. 2024{\natexlab{a}}.
\newblock Faithful Logical Reasoning via Symbolic Chain-of-Thought.
\newblock \emph{arXiv preprint arXiv:2405.18357}.

\bibitem[{Xu et~al.(2024{\natexlab{b}})Xu, Yao, Guo, Cui, Ni, Ge, Chua, Liu, Sun, and Huang}]{xu2024llava}
Xu, R.; Yao, Y.; Guo, Z.; Cui, J.; Ni, Z.; Ge, C.; Chua, T.-S.; Liu, Z.; Sun, M.; and Huang, G. 2024{\natexlab{b}}.
\newblock Llava-uhd: an lmm perceiving any aspect ratio and high-resolution images.
\newblock \emph{arXiv preprint arXiv:2403.11703}.

\bibitem[{Xu et~al.(2020{\natexlab{a}})Xu, Li, Cui, Huang, Wei, and Zhou}]{xu2020layoutlm}
Xu, Y.; Li, M.; Cui, L.; Huang, S.; Wei, F.; and Zhou, M. 2020{\natexlab{a}}.
\newblock Layoutlm: Pre-training of text and layout for document image understanding.
\newblock In \emph{Proceedings of the 26th ACM SIGKDD international conference on knowledge discovery \& data mining}, 1192--1200.

\bibitem[{Xu et~al.(2020{\natexlab{b}})Xu, Xu, Lv, Cui, Wei, Wang, Lu, Florencio, Zhang, Che et~al.}]{xu2020layoutlmv2}
Xu, Y.; Xu, Y.; Lv, T.; Cui, L.; Wei, F.; Wang, G.; Lu, Y.; Florencio, D.; Zhang, C.; Che, W.; et~al. 2020{\natexlab{b}}.
\newblock Layoutlmv2: Multi-modal pre-training for visually-rich document understanding.
\newblock \emph{arXiv preprint arXiv:2012.14740}.

\bibitem[{Yang et~al.(2023)Yang, Li, Wang, Lin, Azarnasab, Ahmed, Liu, Liu, Zeng, and Wang}]{yang2023mm}
Yang, Z.; Li, L.; Wang, J.; Lin, K.; Azarnasab, E.; Ahmed, F.; Liu, Z.; Liu, C.; Zeng, M.; and Wang, L. 2023.
\newblock Mm-react: Prompting chatgpt for multimodal reasoning and action.
\newblock \emph{arXiv preprint arXiv:2303.11381}.

\bibitem[{Ye et~al.(2023)Ye, Hu, Xu, Ye, Yan, Xu, Li, Tian, Qian, Zhang et~al.}]{ye2023ureader}
Ye, J.; Hu, A.; Xu, H.; Ye, Q.; Yan, M.; Xu, G.; Li, C.; Tian, J.; Qian, Q.; Zhang, J.; et~al. 2023.
\newblock Ureader: Universal ocr-free visually-situated language understanding with multimodal large language model.
\newblock \emph{arXiv preprint arXiv:2310.05126}.

\bibitem[{Zeng et~al.(2022)Zeng, Attarian, Ichter, Choromanski, Wong, Welker, Tombari, Purohit, Ryoo, Sindhwani et~al.}]{zeng2022socratic}
Zeng, A.; Attarian, M.; Ichter, B.; Choromanski, K.; Wong, A.; Welker, S.; Tombari, F.; Purohit, A.; Ryoo, M.; Sindhwani, V.; et~al. 2022.
\newblock Socratic models: Composing zero-shot multimodal reasoning with language.
\newblock \emph{arXiv preprint arXiv:2204.00598}.

\bibitem[{Zheng~Cai and Chen(2024)}]{cai2024internlm2}
Zheng~Cai, M.~C.; and Chen, H. 2024.
\newblock InternLM2 Technical Report.
\newblock arXiv:2403.17297.

\end{thebibliography}
\clearpage
\appendix

\section{Generated Data Analysis}
We classify the generated questions into five categories to clarify the model’s performance on different types of questions. Among these, since DocVQA is a scanned document dataset, we did not generate questions involving color information. Additionally, since the training sets of DocVQA and InfoVQA contain a large number of text extraction questions, we limited the types of questions generated by DocVQA to spatial, count, and reasoning when generating incremental questions. InfoVQA limits the types to color, spatial, count, and reasoning in the process of generating incremental questions. For details, as shown in Table \ref{statistic}.
\begin{table}[h]
\resizebox{0.47\textwidth}{!}{
\begin{tabular}{cccccc}
\specialrule{1.5pt}{1.5pt}{1.5pt}
Dataset & Color & Spatial & Text & Count & Reasoning \\ \hline
DocVQA  &0       &3779         &39769      &9386       &5390           \\
InfoVQA &2019       & 4192        & 15820     &10775       & 4026          \\
ChartQA &5094       & 16        &12254      &23596       &26689           \\ \specialrule{1.5pt}{1.5pt}{1.5pt}
\end{tabular}}
\caption{Statistics of different question types of generated data.}
\label{statistic}
\end{table}

\section{Further Analysis of ChartQA}
ChartQA is divided into synthetic data (Augmented) and human-generated data (Human). The questions in Augmented are relatively simple, mostly involving the direct extraction of chart information without complex calculations. Human questions are more complex and include multi-step or nested operations, which better test the model’s reasoning ability. In Table \ref{chartqa}, it is evident that Zero-shot, Few-shot, and Finetune methods perform better than straightforward methods, with the Finetune method we trained being far superior for counting and reasoning questions. For Text\_extractive type data, performance is more unstable, and extracting information from complex charts, including color information, remains a significant challenge. Given the small amount of data in Color and Spatial categories, these results are not definitive but indicate that after fine-tuning, the model’s understanding of color information has declined, which is closely related to the performance of the model’s visual encoder.

\begin{table*}[h]
\centering
\begin{tabular}{ccccccc}
\specialrule{1.5pt}{1.5pt}{1.5pt}
Dataset                    & Strategies & Color                & Text                 & Spatial              & Count                & Reasoning            \\ \hline
\multirow{4}{*}{Augmented} & Straightforward        & -                     & 71.03                     &1.0                      &65.71                      &54.35                      \\
                           & Zero-shot        &-                      & 76.05                     &1.0                      &66.28                      &60.87                      \\
                           & Few-shot        & \multicolumn{1}{c}{-} & \multicolumn{1}{c}{80.37} & \multicolumn{1}{c}{0.0} & \multicolumn{1}{c}{72.91} & \multicolumn{1}{c}{65.22} \\
                           & Finetune        & -                     &80.96                      &1.0                      &81.84                      & 82.61                     \\ \hline
\multirow{4}{*}{Human}     & Straightforward        &67.44                      &81.06                      &0.5                      & 39.27                     &47.13                      \\
                           & Zero-shot        & \multicolumn{1}{c}{65.12} & \multicolumn{1}{c}{79.74} & \multicolumn{1}{c}{0.75} & \multicolumn{1}{c}{42.46} & \multicolumn{1}{c}{49.31} \\
                           & Few-shot        &32.56                      &77.97                      &0.25                      &47.98                      &53.27                      \\
                           & Finetune        &39.53                      &76.65                      &1.0                      &61.15                      &58.61                      \\ \specialrule{1.5pt}{1.5pt}{1.5pt}
\end{tabular}
\caption{ChartQA experimental results on two datasets.}
\label{chartqa}
\end{table*}

\section{Error Analysis}
Although DocAssistant has achieved the most advanced results on the three document datasets, it still has shortcomings. After our analysis, we identified two types of errors.

\begin{figure*}
\centering
\includegraphics[width=0.9\textwidth]{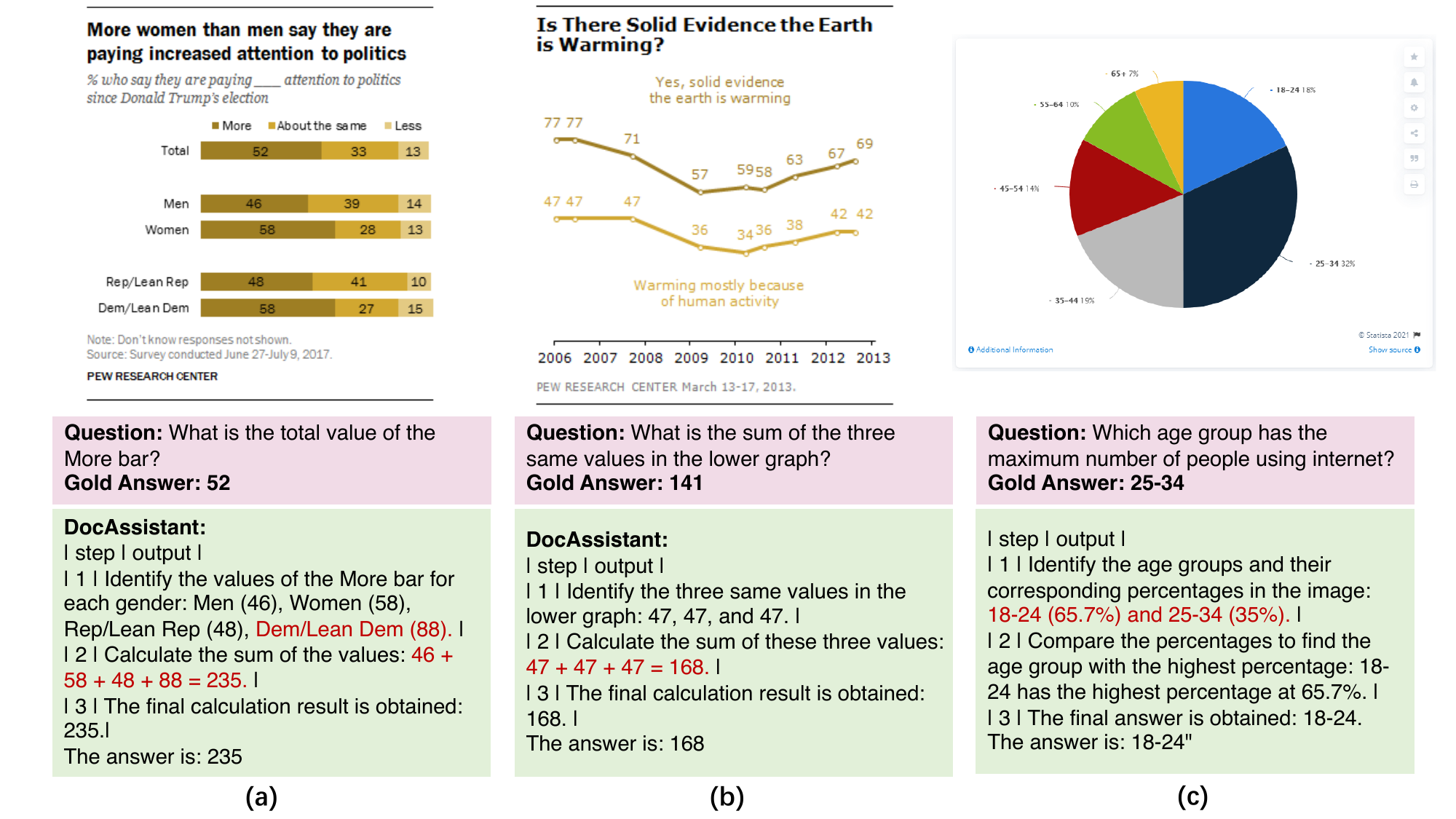}
\caption{DocAssistant error examples on ChartQA.
}
\label{charterror}
\end{figure*}

\begin{figure*}
\centering
\includegraphics[width=0.9\textwidth]{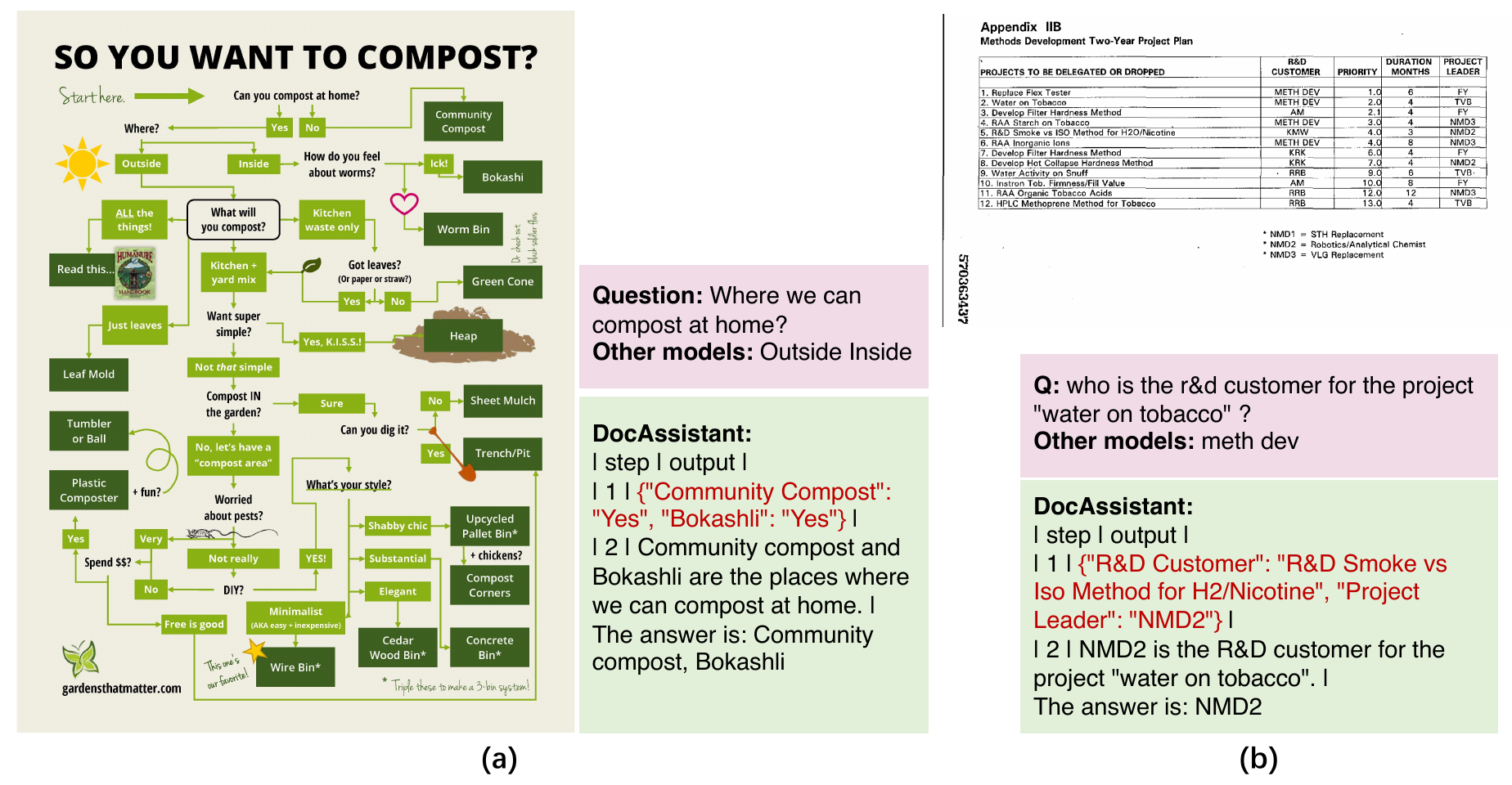}
\caption{DocAssistant error examples on InfoVQA and DocVQA.
}
\label{docerror}
\end{figure*}

\begin{itemize}
    \item One type of error is information recognition. For example, the pie chart in Figure \ref{charterror} (c) fails to identify the percentage corresponding to "18-24". The percentage corresponding to "24-35" is identified incorrectly (32\% is identified as 35\%), and the 7\% corresponding to "65+" is identified as 65.7\%. 
    \item The second type of error is the incorrect correspondence between the problem keyword and the document image information. As shown in Figure \ref{docerror} a, the keyword ‘compost at home’ in the problem incorrectly corresponds to the information ‘Community Compost’ in the image. In Figure \ref{docerror} (b), the keyword "water on tobacco" in the question incorrectly corresponds to "R\&D Smoke vs Iso Method for H2/Nicotine" in the image, creating the illusion that the problem does not appear in the "Project Leader" information. 
    \item The third type of error is the inference error of the intermediate step, as shown in Figure \ref{charterror} (b), which occurs when addition calculations are performed with accurate information identification. 
    \item Finally, all three types of errors may occur simultaneously. As shown in Figure \ref{charterror} (a), no information related to the keywords of the problem is obtained, the information in the chart is identified incorrectly, and there are calculation errors in the intermediate steps.
\end{itemize}

The occurrence of the above errors is related to the performance of the model’s three main modules. Identification errors are mainly due to deficiencies in the visual encoder’s fine-grained understanding of document information. The model’s inability to accurately correspond problem information to image information indicates that the language model still has deficiencies in aligning with visual information. Finally, errors in the intermediate inference step show that the language model is insufficient for inferring and calculating complex information.

\section{Templates}
\subsection{Templates for Generated QA}
Table \ref{r template} shows the instructions for the DocVQA and InfoVQA training sets to generate evidence corresponding to the question using MLLM based on the existing Q\&A pairs. Table \ref{qa template} shows the generation of specified types of questions, corresponding evidence, and answers based on the training set images in DocVQA and InfoVQA. Table \ref{few-shot} shows the few-shot template designed for ChartQA. The design of the few-shot template enables it to flexibly adapt to various questions due to the variety of intermediate steps. Additionally, this template can generate corresponding intermediate reasoning steps and questions for the existing question and answer pairs simultaneously. Table \ref{type} is a list of instructions that constrain questions generated from different datasets.
\subsection{Template for the data checker.}
Since the document may not be suitable for all types of questions, there will be noise and reasoning steps with noise in the generation process. The data checker we designed in Table \ref{checker} operates in two aspects: First, it detects whether there is an information detection error based on the question generation. This detection uses external tools to extract the text information of the image. Second, it checks whether the intermediate reasoning step is correct, such as identifying calculation errors. If an error is detected, the data is discarded. Thus, if either of these two errors is present, the data is discarded to ensure the quality of the generated data.

\subsection{Template for question classification.}
To further analyze which question types DocAssistant has improved, we used a multimodal large language model to classify questions, and the template is shown in Table \ref{classification}. Since existing datasets do not categorize each question specifically, we subjectively designed five categories to cover almost all the questions in the DVQA datasets. We defined each question and provided examples corresponding to each type for different datasets and leveraged the powerful understanding of the MLLM to improve classification accuracy. After our sampling check, the classification of most questions was accurate.
\begin{table*}[h]
\resizebox{0.97\textwidth}{!}{
\begin{tabular}{l}
\specialrule{1.5pt}{1.5pt}{1.5pt}
\begin{tabular}[c]{@{}l@{}}\#\#\# Instruction:\\ \\ Given an image and a question in the following, what is the answer to the question? Please complete the task in two steps:\\ 1. In the first step, extract the relevant contexts related to the keywords in the question from the provided image. \\Store these in the variable "\{evidence\}". If there are multiple contexts, separate them using the "\#" symbol.\\ 2. In the second step, predict the answer based on the \{evidence\} and store it in the variable "\{answer\}".\\ \\ Please organize the results in the following table:\\ | step | output |\\ | 1 | \{evidence\} |\\ | 2 | \{answer\} |\\ \\ The example format of response:\\ \#\#\# Response\\ | step | output |\\ | 1 | \{"SARS": "10\%", "MERS": "34\%", "EBOLA": "50\%+"\} |\\ | 2 | EBOLA has the highest mortality rate. |\\ \\ Follow the format of the instruction above, Generate the corresponding response based on the question and answer. \#\#\#Question\\ question\\ \#\#\#Gold\_answer\\ answer\end{tabular} \\ \specialrule{1.5pt}{1.5pt}{1.5pt}
\\
\end{tabular}}
\caption{DVQA rationale generation template.}
\label{r template}
\end{table*}

\begin{table*}[h]
\resizebox{0.97\textwidth}{!}{
\begin{tabular}{l}
\specialrule{1.5pt}{1.5pt}{1.5pt}
\begin{tabular}[c]{@{}l@{}}\#\#\# Instruction:\\ \\ Given an image in the following, generate a question and the corresponding answer. Please complete the task in three steps: \\1. In the first step, Generate a question, the question should ... Store the question in the Variable \{question\}. \\ 1. In the second step, extract the relevant contexts related to the keywords in the question from the provided image. \\Store these in the variable "\{evidence\}". If there are multiple contexts, separate them using the "\#" symbol.\\ 2. In the third step, predict the answer based on the \{evidence\} and store it in the variable "\{answer\}".\\ \\ Please organize the results in the following table:\\ | step | output |\\ | 1 | \{question\} |\\ | 2 | \{evidence\} |\\ | 2 | \{answer\} |\\ \\ The example format of response:\\ \#\#\# Response\\ | step | output |\\ | 1 | Which disease has the highest mortality rate? |\\ | 2 | \{"SARS": "10\%", "MERS": "34\%", "EBOLA": "50\%+"\} |\\ | 3 | EBOLA has the highest mortality rate. |\\ The answer is: EBOLA. \\ \\ Follow the format of the instruction above, Generate the corresponding response based on the image.\end{tabular} \\ \specialrule{1.5pt}{1.5pt}{1.5pt}
\\
\end{tabular}}
\caption{DVQA question generation template.}
\label{qa template}
\end{table*}

\begin{table*}[h]
\resizebox{0.97\textwidth}{!}{
\begin{tabular}{l}
\specialrule{1.5pt}{1.5pt}{1.5pt}
\begin{tabular}[c]{@{}l@{}}\#\#\#Instruction\\ The following are given a chart image and five examples to complete the task of generating chart question and answer data.\\ \\ Example1:\\     Generate a question and the corresponding answer step by step based on the image:\\     \\     Question: What is the difference in value between Lamb and Corn?\\     Answer: \\         Step1. The values of relevant indicators in the question are identified: The value of Lamb is 103.7 and the value of Corn is 103.13,\\         Step2. Perform the calculation of the difference in value between Lamb and Corn: 103.7-103.13=0.57,\\         Step3. The final calculation result is obtained: 0.57.\\ \\ Example2:\\     Generate a question and the corresponding answer step by step based on the image:\\ \\     Question: In which year is the difference between the green and blue graphs lowest?\\     Answer: \\         Step1. Identify the years of the chart: 2017, 2018, 2019.,\\         Step2. Compare the values of the green and blue graphs for each year: in 2017, the green graph is at 65 and the blue graph is at 56.\\ In 2018, the green graph is at 70 and the blue graph is at 41. In 2019, the green graph is at 69 and the blue graph is at 50.,\\         Step3. Calculate the difference between the green and blue graphs for each year: In 2017: 65-56=9, in 2018: 70-41=29, in 2019: 69-50=19,\\         Step4. Sort all the differences": "In 2017: 9, in 2018: 19, in 2019: 29,\\         Step5. Perform the calculations required in the question: in 2017, the difference between green and blue graphs is the lowest,\\         Step6. The final calculation result is obtained: 2017.\\ \\ Example3:\\     Generate a question and the corresponding answer step by step based on the image:\\ \\     Question: What's the average of all the values in the green bars?\\     Answer: \\         Step1. Identify all information of the blue bar: \{"Characteristic": "US, EU, China", "More": "29, 19, 17"\},\\         Step2. Perform the calculation of the average of all the values in the green bars: 29 + 19 + 17 = 21.6,\\         Step3. The final calculation result is obtained: 21.6.\\ \\ Example4:     \\     Generate a question and the corresponding answer step by step based on the image:\\ \\     Question: Which country has the third highest rate of cases in Europe?\\     Answer: \\         Step1. Identify all values of countries in Europe: \{"Montenegro":16111.01, "Czechia": 15 587.77, "Sweden": 10546.7, "Slovenia": 12276, "Slovakia":14259.69\},\\         Step2. Sort all values: \{"Montenegro":16111.01, "Czechia": 15 587.77, "Slovakia":14259.69, "Slovenia": 12276, "Sweden": 10546.7\},\\         Step3. The third highest rate of cases in Europe is obtained: Slovakia.\\ \\ Example5:\\     Generate a question and the corresponding answer step by step based on the image:\\ \\     question: What is the sum of all the blue bar?\\     Answer: \\         Step1. Identify all information of the blue bar: \{"Characteristic": "Number of gamers in millions", "2012": 8.12, "2013": 9.04, "2014": 9.97\},\\         Step2. Calculate the sum of all values: 8.12+9.04+9.97=27.13,\\         Step3. The final calculation result is obtained: 27.13.\\ \\ Follow the format of the example above, generate a question and the corresponding answer step by step based on the image.\end{tabular} \\ \specialrule{1.5pt}{1.5pt}{1.5pt}
\end{tabular}}
\caption{Chart question generation or rationale generation from the few-shot template.}
\label{few-shot}
\end{table*}

\begin{table*}[h]
\resizebox{0.97\textwidth}{!}{
\begin{tabular}{l}
\specialrule{1.5pt}{1.5pt}{1.5pt}
\begin{tabular}[c]{@{}l@{}}DocVQA \& InfoVQA \& ChartQA\\ 1. The question should require spatial understanding of the image.\\ 2. The question should require counting.\\ 3. The question should require reasoning of the image.\\ \\ InfoVQA \& ChartQA\\ 1. The question should require color understanding of the image.\\ 2. The question should require counting of colors.\\ 3. The question should require counting and color understanding.\\ \\ ChartQA\\ 1. The question should require math reasoning about min.\\ 2. The question should require math reasoning about average.\\ 3. The question should require math reasoning about the difference between max and min.\\ 4. The question should require math reasoning about difference.\\ 5. The question should require math reasoning about comparison.\\ 6. The question should require math reasoning about average and max.\\ 7. The question should require math reasoning about sum.\\ 8. The question should require math reasoning about max.\\ 9. The question should require math reasoning about average and min.\\ 10. The question should require math reasoning about ratio.\\ 11. The question should require color understanding and math reasoning to compute the difference.\\ 12. The question should require color understanding and math reasoning about comparison.\\ 13. The question should require spatial understanding and math reasoning to compute difference.\\ 14. The question should require spatial understanding and math reasoning about average.\end{tabular} \\ \hline
\end{tabular}}
\caption{Constraints on question generation for different datasets.}
\label{type}
\end{table*}

\begin{table*}[h]
\resizebox{0.97\textwidth}{!}{
\begin{tabular}{l}
\hline
\begin{tabular}[c]{@{}l@{}}Below is an instruction that describes an evidence error detection task in the general document domain, paired with an image.\\ Generate an appropriate response to the given instruction.\\ \\ \#\#\# Instruction:\\         Given an image, question-answer pair, and corresponding evidence, assess whether the evidence is faithful to the images and \\corresponding text information, and whether it accurately contains the context information of the question-answer pair in the\\ image and table. Please complete the task in three steps:\\         1. In the first step, assess whether each step of evidence is consistent with the information in the image and the table. \\If consistent, store "True" in the variable \{is\_faithful\}; otherwise, store "False".\\         2. In the second step, assess whether each step of evidence contains the context information of the question-answer pair in the image.\\ If it does, store "True" in the variable \{is\_include\}; otherwise, store "False".\\ 	3. If \{is\_faithful\} is True and \{is\_include\} is True, store "True" in the variable \{result\}; otherwise, store "False" in the variable \{result\}.\\         Please organize the results in the following table:\\         | step | output |\\         | 1 | \{is\_faithful\} |\\         | 2 | \{is\_include\} |\\         Finally, present the predicted answer in the format: "The answer is: \{result\}"\\ \\         \#\#\#Follow the example:\\ \\         \#\#\# Question\_answer pairs\\         LIver is a source of how many of the vitamins shown here? answer: 6\\         \#\#\# Evidences\\         | step | output |\\         | 1 | 40\% of visitors to VIC went to Melbourne. |\\         | 2 | 60\% |\\         The answer is: 60\%\\         \\         \#\#\# Response\\         |step | output|\\         |1 | False |\\         |2 | True |\\         The answer is: False\\ \\ Follow the format of the instruction above, Generate the corresponding response based on the image, evidence, and question-answer pairs:\\ \#\#\#Question\_answer pairs\\ question \& answer\\ \#\#\#Evidence\\ rationale\end{tabular} \\ \specialrule{1.5pt}{1.5pt}{1.5pt}
\end{tabular}}
\caption{Error detection template for generated data.}
\label{checker}
\end{table*}

\begin{table*}[h]
\resizebox{0.97\textwidth}{!}{
\begin{tabular}{l}
\specialrule{1.5pt}{1.5pt}{1.5pt}
\begin{tabular}[c]{@{}l@{}}Given a dataset consisting of DocVQA/InfographicsVQA/ChartQA and corresponding questions. \\The task is to classify each question into one of the following five types based on the image and the type of information\\ required to answer the question.\\ \\         Here are the definitions for each type:\\ \\         Color: Questions that require an understanding of colors.\\         Spatial: Questions that involve spatial relationships or positions (e.g., "next to," "above," "below", "left", "right").\\         Text\_extractive: Questions that require extracting specific text information from the document image.\\         Count: Questions that involve counting elements or objects in the document image.\\         Reasoning: Questions that require logical reasoning, inference, or combining multiple pieces of information.\\ \\         For each question, analyze the content and determine the appropriate type. \\Now, classify the following questions from the DocVQA/InfographicVQA/ChartQA dataset:\\ \#\#\#Question\\     question\end{tabular} \\ \specialrule{1.5pt}{1.5pt}{1.5pt}
\end{tabular}}
\caption{Question classification template based on MLLM.}
\label{classification}
\end{table*}

\end{document}